\documentclass[preprint,aps,prd,numbers,sort&compress,nofootinbib,mcite]{revtex4-1}
\usepackage{graphicx}
\usepackage{float}
\usepackage{bm}
\usepackage{amssymb}
\usepackage{slashed}
\usepackage{amsmath}
\usepackage{verbatim}
\usepackage{color}
\usepackage{xcolor}
\usepackage{subfigure}
\usepackage{epstopdf}
\usepackage{multirow}
\usepackage{tabularx}

\usepackage{hyperref}
\hypersetup{colorlinks=true,citecolor=blue,urlcolor=blue,linkcolor=blue}

\newcounter{JW}

\begin{document}
\title{ Searching for the charged-current non-standard neutrino interactions at the $e^{-}p$ colliders}

\author{Chong-Xing Yue}
\email{cxyue@lnnu.edu.cn}
\author{Xue-Jia Cheng}
\email{cxj225588@163.com}
\author{Yue-Qi Wang}
\email{wyq13889702166@163.com}
\author{Yan-Yu Li}
\email{lyy3390@163.com}
\date{\today}
\affiliation{Department of Physics, Liaoning Normal University, Dalian 116029, China}

\begin{abstract}
Considering the theoretical constraints on the charged-current~(CC) non-standard neutrino interaction~(NSI) parameters in a simplified $W'$ model, we study the sensitivities of the Large Hadron electron Collider~(LHeC) and Future Circular Collider-hadron electron~(FCC-he) to the CC NSI parameters through the subprocess $e^{-}q \to W' \to v_{e}q'$. Our results show that the LHeC with $\mathcal{L} = 100\;{\rm fb}^{-1}$ is a little more sensitive to the CC NSI parameters than that of the high-luminosity~(HL)~LHC, and the absolute values of the lower limits on the CC NSI parameters at the FCC-he are smaller than those of the HL-LHC about two orders of magnitude.
\end{abstract}

\maketitle

\section{\label{level1} Introduction }
The Standard Model~(SM) has been a powerful method for studying particle physics. However, the problem of neutrino oscillations~\cite{Super-Kamiokande:1998kpq} arising from mismatching flavor and mass eigenstates is one of the open questions that seems to be unexplainable within the traditional framework of the SM, highlighting the need for new physical theories that go beyond the SM~(BSM)~\cite{Barger:2003qi,Gonzalez-Garcia:2007dlo}.
With the concentration of efforts in this area, the study of non-standard neutrino interactions~(NSIs) is gradually emerging, as detailed in Refs.~\cite{Wolfenstein:1977ue,Ohlsson:2012kf,Miranda:2015dra,Choudhury:2018xsm}.
Many studies have constrained NSIs (both charged-current and neutral-current) to be of the order of $\mathcal{O}(10^{-4})$ to $\mathcal{O}(10^{-1})$, depending on the data used for the analysis and the model chosen~\cite{Santos:2020dgs,Davidson:2011kr,Gonzalez-Garcia:2011vlg,Gonzalez-Garcia:2013usa,Proceedings:2019qno,Liao:2017awz,Bakhti:2016gic}.
 NSIs are universal methods for studying the effects of new physics in neutrino oscillations. The neutrino production and detection processes involving non-standard interactions are usually associated with charged leptons and are referred to as charged current NSIs~(CC NSIs). The experiments usually impose stricter constraints on the CC NSIs than those for the neutral-current NSIs~(NC NSIs)~\cite{Davidson:2003ha,Biggio:2009nt}.

The presence of NSIs not only has a great impact on the precision measurements of neutrino oscillation experiments, but also is an important platform for studying the impact of new physics in high- and low-energy collider experiments and leads to a very rich phenomenology~\cite{Yue:2021snv,Choudhury:2018xsm,Gonzalez-Garcia:2011vlg,Gonzalez-Garcia:2013usa,Davidson:2011kr,Yue:2020wkj,Babu:2019mfe,Terol-Calvo:2019vck,Altmannshofer:2018xyo,Cirigliano:2012ab,Santos:2020dgs,Liao:2017awz,Proceedings:2019qno,Bakhti:2016gic}.
In particular, the effect of the CC NSI in neutrino oscillation experiments depends on the strength of the new vector interaction, which strongly depends on the flavor structure of the CC NSI, leading to a degeneracy that is difficult to break by neutrino facilities~\cite{Babu:2020nna}. However, the collider data can break the degeneracy because of its insensitivity to neutrino flavor and the signal will lead to the same observation for different flavors.
So studying NSIs at the collider experiments are used as a complement to the neutrino oscillation experiments.

The new particles and the new interaction processes predicted by the new physics models are able to produce a wealth of physical phenomena in the high-energy scales, and it is entirely possible that the already operating and future high-energy colliders will detect these new physical signals in the future. In special, the new charged gauge boson $W'$, a hypothetical heavy partner of the SM $W$ gauge boson, is an important detection goal of future colliders, which can contribute to the CC NSI parameter as $\epsilon \sim g^{2}_{W'} / M^{2}_{W'} $ in a simplified model framework.  In this paper we will study the sensitivities of the $e^{-}p$ colliders  to the CC NSI parameters via the subprocess $e^{-}q \to W' \to v_{e} q'$.

Currently, in order to combine the excellent performance of proton-proton collider and electron-electron collider, the $e^{-}p$ collider has been proposed, which can complement the proton ring with an electron beam and allow deep inelastic leptonic scattering~(DIS) of electrons and protons at TeV energies. Its kinematics extends to higher scales relative to HERA~\cite{LHeC:2020van,LHeCStudyGroup:2012zhm,AbelleiraFernandez:2012ni,Antusch:2020fyz,Bruening:2013bga,FCC:2018byv,Spor:2021dhf,Gutierrez-Rodriguez:2020gsi}.
Compared to the $e^{+}e^{-}$ collider, the $e^{-}p$ collider allows DIS studies of the internal structure of nuclei. At the same time it is also more advantageous compared to the $pp$ collider. The $e^{-}p$ collider can provide a cleaner background that suppresses the background of strong QCD interactions, providing high precision measurements. In addition, since the initial state is asymmetric, the backward and forward scattering can be disentangled, which can greatly increase the significance of the signal and get opportunities that cannot be observed in $pp$ collider. Thus, the $e^{-}p$ collider not only becomes a good choice to complement the $pp$ and $e^{+}e^{-}$ colliders, but also provides distinctive ways to precision Higgs physics, top quark, electroweak physics as well as new physics beyond the SM~\cite{Hesari:2018ssq}.

Two major future colliders have been designed at CERN for deeply inelastic lepton-hadron scattering: the Large Hadron electron Collider~(LHeC)~\cite{LHeC:2020van,LHeCStudyGroup:2012zhm,Antusch:2020fyz,AbelleiraFernandez:2012ni,Bruening:2013bga} and the Future Circular Collider-hadron electron~(FCC-he)~\cite{FCC:2018byv,Spor:2021dhf,Gutierrez-Rodriguez:2020gsi}, which aim to combine the proton beams of the Large Hadron Collider~(LHC) with a new electron accelerator at the main tunnel of the LHC. The values of the center-of-mass (c.m.) energy $\sqrt{s}$ and integrated luminosity $\mathcal{L}$ of the LHeC and FCC-he are specified in Table~\ref{Tab:ep}.
\begin{table}[H]
	\centering
	\caption{\label{Tab:ep}The designed values of the c.m. energy $\sqrt{s}$ and integrated luminosity $\mathcal{L}$ for the $e^{-}p$ colliders. }
	\label{table1}
	 \begin{tabular}{|p{2.5cm}<{\centering}|p{2cm}<{\centering}|p{2cm}<{\centering}|p{2cm}<{\centering}|p{4cm}<{\centering}|}
		\hline
        Colliders  &$\sqrt{s}~({\rm TeV})$ &$E_{e}~({\rm GeV})$&$E_{p}~({\rm TeV})$&$\mathcal{L}~({\rm fb}^{-1})$\\
        \hline
		\multirow{2}{*}{LHeC}	&  1.30   	&  60 & 7 &\multirow{2}{*}{ 10, 30, 50, 100 }\\ \cline{2-4}
		   &   1.98   & 140&7 &  \\
		\hline
		\multirow{2}{*}{FCC-he}	&	3.46  & 60& 50&\multirow{2}{*}{100, 500, 1000, 2000} \\ \cline{2-4}
       	&	5.29  &140 & 50&\\
		\hline	
	\end{tabular}
\end{table}
In this paper, considering the theoretical constraints on the CC NSI parameters in a simplified $W'$ model, we analysis the signal and the relevant background of the subprocess $e^{-}q \to v_{e}q'$ including the CC NSI contributions via Monte-Carlo~(MC) simulations and discuss the sensitivities of the LHeC and FCC-he to the CC NSI parameters. In our numerical calculations, we consider the contributions of the interference term between the CC NSI and the SM. Our results show that the sensitivity of the LHeC to the CC NSI parameters is in the same order of magnitude as that from the high-luminosity~(HL)~LHC, on the other hand, the sensitivity of the FCC-he to the CC NSI parameters is more sensitive.

Our paper is organised as follows. In Sec.~\ref{level2}, we review the theoretical constraints on the CC NSI parameters $\epsilon^{qq'Y}_{\alpha\beta}$ in a simplified $W'$ model.
The sensitivities of the LHeC and FCC-he to the CC NSI parameters $\epsilon^{qq'L}_{\alpha\beta}$ via the subprocess $e^{-}q \to v_{e}q'$ are studied in Sec.~\ref{level3}. Our conclusion is given in Sec.~\ref{level5}.

\section{\label{level2}Theoretical model framework }
 NSIs are usually described by six-dimensional four-fermion operators~\cite{Proceedings:2019qno,Biggio:2009nt,Biggio:2009kv,Davidson:2003ha}, where for CC NSIs with quarks are given by effective Lagrangian
\begin{equation}
\mathcal{L}_{NSI,CC} =  -2{\sqrt{2}}{G_{F}}\epsilon^{qq'Y}_{\alpha\beta}\left[q\gamma^{\mu}P_{Y}q'\right]\left[\bar{\nu_{\alpha}}\gamma_{\mu}P_{L}\ell_{\beta}\right]+h.c.
\end{equation}
Here ${G_{F}}$ is the Fermi constant and $P_{Y}$ is a chiral projection operator including $P_{L}$ or $P_{R}$.
The parameters $\epsilon^{qq'Y}_{\alpha\beta}$ ($Y = L$ or $R$) are dimensionless coefficients that quantify the strengthes of the new vector interactions. We have assumed that there are only left-handed neutrinos in above equation.

The constraints on the CC NSI parameters have been studied at the LHC~\cite{Yue:2021snv,Cirigliano:2012ab}, but relevant study at the $e^{-}p$ collider is currently not available. In this paper, we mainly study the possibility of detecting the CC NSI parameters induced by exchange of the new gauge boson $W'$ at the LHeC and FCC-he. Since the momentum transfers can be high to resolve further dynamics of the new physics in the collider, the impact of the CC NSI may not be simply described by the effective operators. In this paper, we focus on a simplified model with the CC NSI induced by exchange of a $W'$ boson. The effective Lagrangian can be written as~\cite{Sullivan:2002jt,Fuks:2017vtl},
\begin{equation}
\mathcal{L}_{W'} =
-\frac{g}{\sqrt{2}}\left[V_{qq'}\bar{q}\gamma^{\mu}\left({A}^{qq'}_{L}P_{L}+{A}^{qq'}_{R}P_{R}\right)q'+{B}^{\alpha\beta}_{L}\bar{\ell_{\alpha}}\gamma^{\mu}P_{L}\nu_{\beta}\right]{W}^{'}_{\mu}+h.c.,
\label{eq.antsaz2}
\end{equation}
where ${W}^{'}_{\mu}$ denotes the new force carrier with mass ${M}_{W'}$, $g$ is the electroweak coupling constant, and $V_{qq'}$ is the Cabbibo-Kobayashi-Maskawa~(CKM) matrix element. $q$ and $q'$ are up-type and down-type quarks.
As long as the square of momentum transfer is much smaller than $M^{2}_{W'}$, the effective parameters $\epsilon^{qq'Y}_{\alpha\beta}$ can be approximatively written as
\begin{equation}
\epsilon^{qq'Y}_{\alpha\beta} =
{A}^{qq'}_{Y} {B}^{\beta\alpha\ast}_{L}\left(\frac{M_{W}}{M_{W'}}\right)^{2}.
\label{eq.antsaz3}
\end{equation}
For the subprocess $e^{-}q \to v_{e}q'$ considered in this paper, there is  $\alpha = \beta = e$. If we assume that this process is flavor conservation, there are only two combinations, i.e. the $ud$ and $cs$ quarks.
 Ref.~\cite{Fuentes-Martin:2020lea} has shown that the luminosity of the $ud$ quarks is about two orders of magnitude higher than that of the $cs$ quarks. If $\epsilon ^{cs Y}_{ee}$ was at the same magnitude as $\epsilon ^{ud Y}_{ee}$, the contribution of $c e^{-}\to \nu_{e} s$ is negligible, therefore we neglect the contribution from a none zero $\epsilon ^{cs Y}_{ee}$.
In the following, $\epsilon ^{ud Y}_{ee}$ is simplified as $\epsilon ^{Y}$.

Ref.~\cite{Yue:2021snv} derives theoretical constraints on the CC NSI parameters from perturbative unitarity and $W'$ decays.
For $M_{W'} \leq 4\;{\rm TeV}$ and $\Gamma _{W'}/ M_{W'} = 0.1$, the $W'$ decays can generate  more strongly constraint as $\left|\epsilon^{L }\right|\leq 10.868\times M_W^2/M_{W'}^2$. For larger $M_{W'}$, the perturbative unitarity constraints are stronger than that from $W'$ decays, specifically, there are $\left|\epsilon^{L }\right|\leq 7.584\times M_W^2/M_{W'}^2$, $\left|\epsilon^{L}\right|\leq 3.637\times M_W^2/M_{W'}^2$ and $\left|\epsilon^{L}\right|\leq 2.089\times M_W^2/M_{W'}^2$ for $M_{W'} = 5\;{\rm TeV}$, $6\;{\rm TeV}$ and $7\;{\rm TeV}$, rspectively.
In the following MC analysis, we will also consider the theoretical constraints on the CC NSI parameters.

\section{\label{level3} The sensitivities of the $e^{-}p$ colliders to the CC NSI parameters  }
Compared to the $pp$ collider, the $e^{-}p$ collider provides an environment in which new physics can be probed with relatively low background rates.
In the post-LHC era, the future $e^{-}p$ colliders such as the LHeC and FCC-he will study the properties of SM observables with unprecedented precision and will be able to discover new particles by direct detection.

So far, little research has been done on the possibility of detecting the CC NSI at the $e^{-}p$ colliders. In this paper, the signal of the process $e^{-}p\to v_{e}j$ including the CC NSI contributions at the LHeC and FCC-he will be calculated and analyzed. In the next analytical calculations, the LHeC is assumed to utilize $7\;{\rm TeV}$ proton beam from the LHC and a $60\;{\rm GeV}$ electron beam with the integrated luminosity as $10\;{\rm fb}^{-1}$ and $100\;{\rm fb}^{-1}$. For the FCC-he, its c.m. energy $\sqrt{s}$ and integrated luminosity $\mathcal{L}$ are taken as $3.46\;{\rm TeV}$ and $100\;{\rm fb}^{-1}$, respectively.

The effective Lagrangian Eq.(\ref{eq.antsaz2}) is implemented in the \verb"FeynRule" program~\cite{Alloul:2013bka,Christensen:2008py,Degrande:2011ua} and then the Universal FeynRules Output~(UFO)
model is inserted to \verb"MadGraph5_aMC@NLO" toolkit~\cite{Alwall:2014hca} and MC simulations are performed.
The renormalization scale $\mu_r$ and factorization scale $\mu_f$ are chosen to be dynamical which are set event-by-event as $(\prod^n_i(M^2_i + p^{i}_{T}))^{\frac{1}{n}}$, with $i$ running over all heavy particles. The events are then showered by \verb"PYTHIA8"~\cite{Sjostrand:2014zea}.
A fast detector simulation is applied by using \verb"Delphes"~\cite{deFavereau:2013fsa} with the LHeC and FCC-he detector cards.
The parton distribution functions are taken as the \verb"NNPDF2.3"~\cite{Ball:2011mu,Ball:2012cx}.
It is well known that the polarization of the primary state electrons in the $ ep$ collider can affect the size of the production cross-section.
After comparison we find that using the beam polarization as $P(e^{-}) = -90\%$ can maximize the cross-section. In the rest of the calculations, the electron beam polarization is taken as $P(e^{-}) = -90\%$,  the basic cuts are selected as $p_{T}^{j}> 20\;{\rm GeV}$ and $ |\eta_{j}| < 5$, and the parameter $\epsilon^{Y}$ is assumed as a real number.
For $\epsilon ^{L}\neq 0$ or $\epsilon ^{R}\neq 0$, the production cross-section of the process $e^{-}p \to v_{e}j $ including the $W'$ contributions can be parameterized as
\begin{equation}
\sigma_{Tot}(\epsilon^{Y})= \sigma_{SM} + \sigma_{INT}(\epsilon^{L}) + \sigma_{NSI}(\epsilon^{Y}).
\label{eq.antsaz7}
\end{equation}
$\sigma_{SM}= 414.12 \;{\rm pb}$ at the $1.30\;{\rm TeV}$ LHeC and $\sigma_{SM}= 936.57 \;{\rm pb}$ at the $3.46\;{\rm TeV}$ FCC-he are the SM cross-sections, $\sigma_{INT}(\epsilon^{L}) = \alpha_{int} \times \epsilon^{L}$ is the interference term between the SM and $W'$ contributions, and $\sigma_{NSI}$ represents the contribution only from $W'$ exchanges with $\sigma_{NSI}(\epsilon^Y) = \alpha^Y_{nsi} \times (\epsilon^Y)^{2}$.
The dependencies of the factors $\alpha_{int}$ and $\alpha^Y_{nsi}$ on $M_{W'}$ can be fitted with the results of the MC simulation for different $M_{W'}$ at the LHeC and FCC-he, which are listed in Table~\ref{fitted}. We find that the interference term plays an important role  and its weight of the total cross-section  can be compared to the measurement accuracy of the LHeC or FCC-he. The sensitivities of the LHeC and FCC-he to the CC NSI parameters have reached the region where interference effect should be considered.
\begin{table}[H]
	\centering
	\caption{\label{fitted}The factors $\alpha_{int}$ and $\alpha^Y_{nsi}$ fitted for different $M_{W'}$ at the LHeC and FCC-he.}
	\label{table1}	 \begin{tabular}{|p{1.8cm}<{\centering}|p{1.5cm}<{\centering}|p{1.5cm}<{\centering}|p{1.5cm}<{\centering}|p{1.5cm}<{\centering}|p{1.5cm}<{\centering}|p{1.5cm}<{\centering}|}
		\hline
        \multicolumn{7}{|c|}{$e^{-}p \to v_{e}j $}\\
        \cline{1-7}
		\multirow{2}{*}{$M_{W'}$(TeV)} & \multicolumn{3}{c|}{LHeC}&\multicolumn{3}{c|}{FCC-he}\\
		\cline{2-7}		
		& $\alpha_{int}$(pb) &$ \alpha_{nsi}^{L}$(pb) &$ \alpha_{nsi}^{R}$(pb) & $\alpha_{int}$(pb) & $\alpha_{nsi}^{L}$(pb)&$\alpha_{nsi}^{R}$(pb)\\
		\hline
		1	& 1473.68    	& 4211.53 &2232.87	&3276.63   &15979.55 &10619.59  \\
		\hline
		2  & 1509.42     & 5000.59 & 2449.67 & 3422.03  &25791.97   &14661.44\\
		\hline
		3	&1517.84	& 5239.23&2493.37	&3473.48	&28376.92	&15956.36\\
		\hline
		4    &1518.76	&5269.97 &2511.75	&3492.44	&32964.58  &16564.58\\
		\hline
		5	&1521.99	&5310.08 &2521.12	&3501.69	&33421.66	&16959.01\\
		\hline
		6	&1523.48	&5316.49 &2524.99	&3508.29	&35148.27	&17238.02\\
		\hline
		7	&1524.22	&5334.72 &2527.39	&3509.89	&35634.52	&17299.33\\
		\hline
	\end{tabular}
\end{table}

In order to distinguish signals from all relevant backgrounds, the \verb"MadAnalysis5"~\cite{Conte:2012fm} was used to analyze kinematic cuts of reconstruction-level events of the signals and backgrounds.
The process $e^{-}p\to v_{e}j$ is mediated by the exchange of $W$ and $W'$ in the t-channel, according to $\mathcal{M} \varpropto 1/(t-M_{W,W'}^2)$, $\mathcal{M}$ is inversely proportional to $(t-M_{W,W'}^2)$, so when $|t|$ becomes small, the amplitude is increased. The events are distributed in the region of large amplitude, so most of the background and signal events will be distributed in the region of small $|t|$. However, in comparison, $\mathcal{M}_{SM}$ will increase much more than the $\mathcal{M}_{W'}$, so a larger proportion of the background events will be distributed in the region of small $|t|$. In order to highlight the signal, the region of small $|t|$ should be cut off. The smaller the transverse momenta of jets $p^{j}_{T}$, the smaller $|t|$ will be, so regions with smaller $p^{j}_{T}$ should be cut off.
For the same reason, the most important variables that distinguish signals from background are also the missing transverse energy $\slashed{E}_T$ and the visible transverse hadronic energy $H_{T}$. Therefore, the kinematic cuts chosen in this paper are $p^{j}_{T}$, $\slashed{E}_T$ and $H_{T}$.
For simplicity of illustration, we show in Fig.~\ref{fig:1} the normalized distributions of $p^{j}_{T}$, $\slashed{E}_T$ and $H_{T}$ for the signals and their associated backgrounds at the LHeC.
\begin{figure}
	\centering
	\subfigure[]{	
\centering	
			\begin{minipage}[t]{0.49\linewidth}
			\includegraphics [width=10cm] {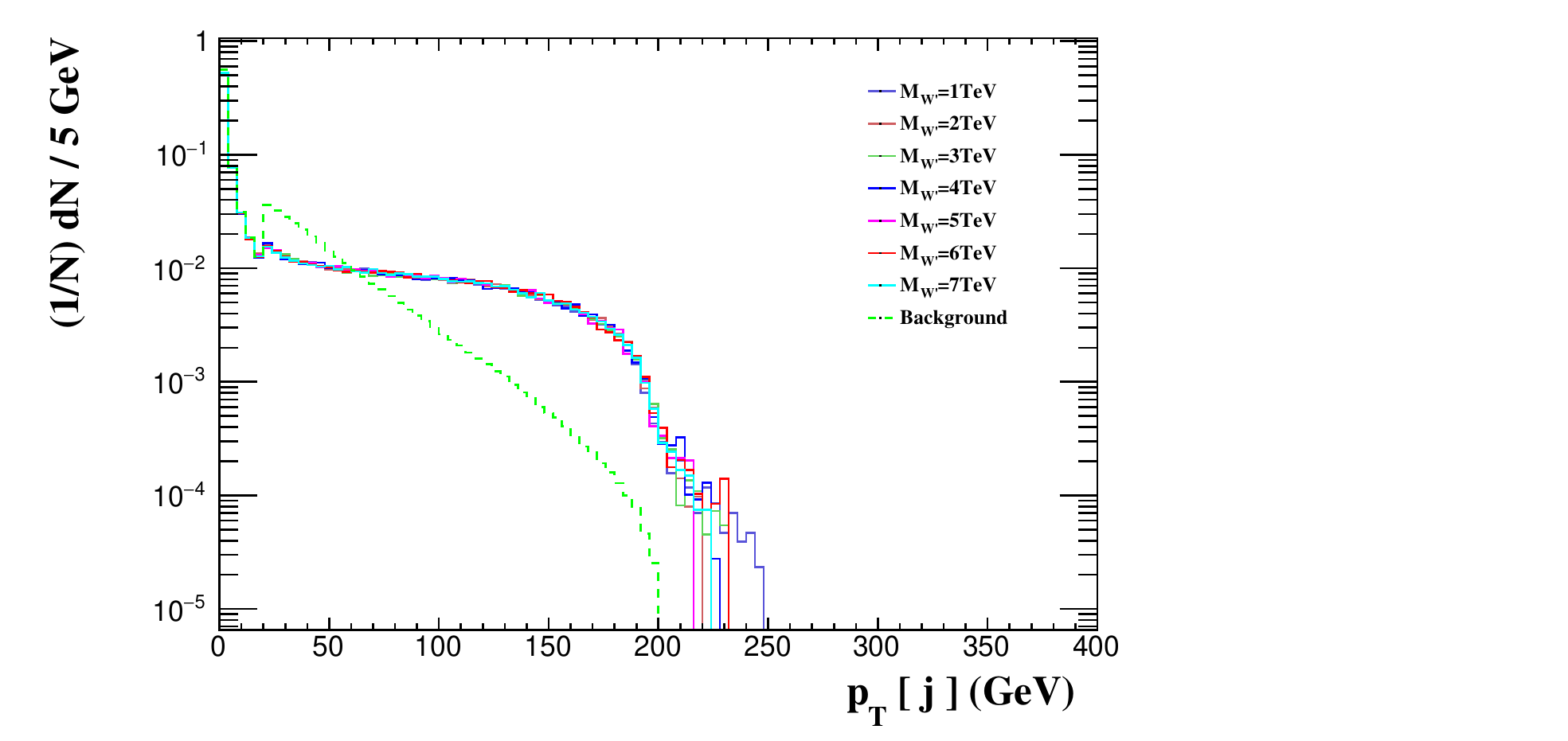}
		\end{minipage}%
	}%
\subfigure[]{	
\begin{minipage}[t]{0.49\linewidth}
	\includegraphics [width=10cm] {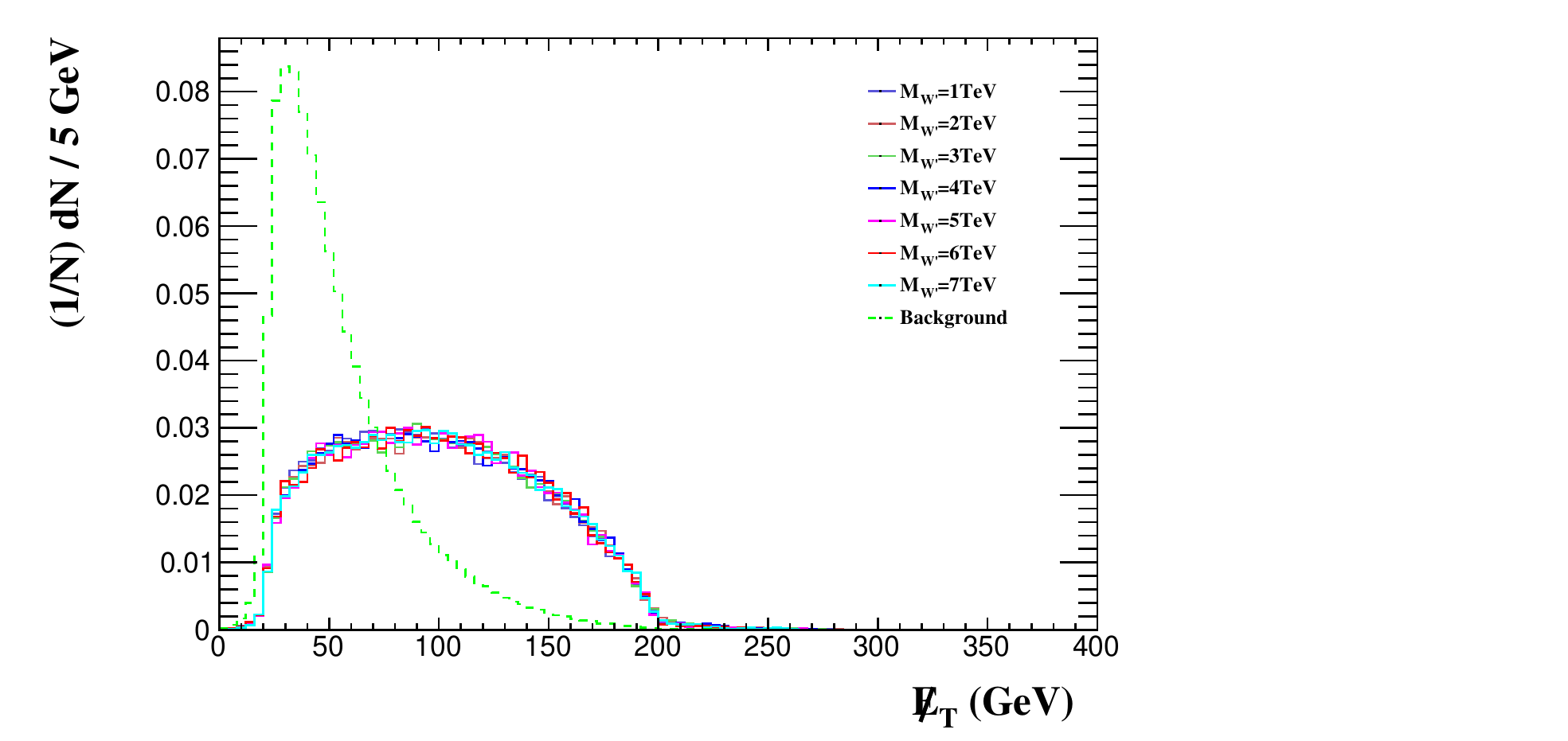}
\end{minipage}%
}%
\quad
\subfigure[]{	
\begin{minipage}[t]{0.49\linewidth}
	\includegraphics [width=10cm] {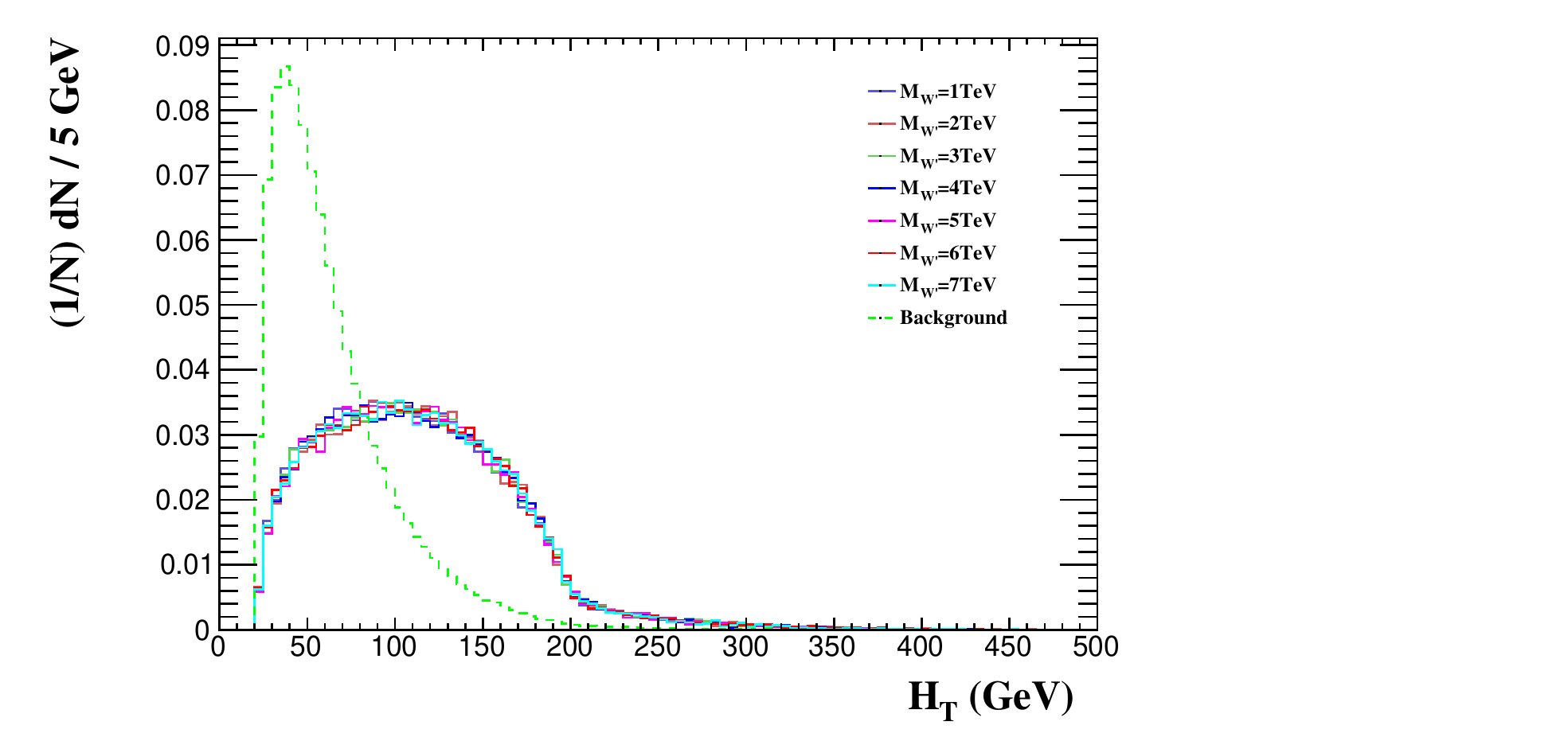}
\end{minipage}%
}%
\caption{\label{fig:1}For the process $e^{-}p \to v_{e}j$, the normalized distributions of $p_{T}^{j}$, ${E\mkern-10.5mu/}_{T}$ and $H_{T}$ for the signals and backgrounds at the LHeC. }
\label{fig:1}
\end{figure}
It can be seen that the signals have distinct characteristic behaviors compared to the associated background. To trigger the signals, we need to further select the cuts according to the associated kinematic distribution, which help to suppress the SM background more effectively.
Since $M_{W'}$ is much heavier than $M_{W}$, the dependence of its production cross-section on the mass $M_{W'}$ is relatively weak. Therefore, the kinematic distribution of the signals does not significantly change with $M_{W'}$.
The improved cuts are shown in Table~\ref{Tab:experiments}.

\begin{table}[H]
	\centering
	\caption{\label{Tab:experiments}Selected kinematic cuts at the LHeC and FCC-he. }
	\label{table2}
	 \begin{tabular}{|p{8cm}<{\centering}|p{3cm}<{\centering}|p{3cm}<{\centering}|}
		\hline
		\multirow{1}{*}{Cuts}  & \multicolumn{1}{c|}{LHeC}&\multicolumn{1}{c|}{FCC-he}\\
		
		\hline
		Cut-1: the transverse momenta of jets 	& $p^{j}_T \geq 80 \;{\rm GeV}$    	&$p^{j}_T \geq 110\;{\rm GeV}$    \\
		\hline
		Cut-2: the missing transverse energy &  $\slashed{E}_T \geq 130\;{\rm GeV}$   &$\slashed{E}_T \geq 150\;{\rm GeV}$   \\
		\hline
		Cut-3: the visible transverse hadronic energy	&$H_{T} \geq 170\;{\rm GeV}$	&$H_{T} \geq 160\;{\rm GeV}$ \\
		\hline	
	\end{tabular}
\end{table}

After the improved cuts are applied, in Tables~\ref{Tab:signal1} and~\ref{Tab:signal2}, we summarize the cross sections of the signal and background at the $\sqrt{s} = 1.30\;{\rm TeV}$ LHeC and $\sqrt{s} = 3.46\;{\rm TeV}$ FCC-he, respectively. As can be seen from these tables, the background is strongly suppressed, while the signal still has good efficiency after all cuts are applied.

\begin{table}[H]
	\centering
	\caption{\label{Tab:signal1} After different cuts applied, the cross sections for the signal and SM background at the $\sqrt{s} = 1.30\;{\rm TeV}$ LHeC. }
	\label{table3}	 \begin{tabular}{|p{2cm}<{\centering}|p{1.5cm}<{\centering}|p{1.5cm}<{\centering}|p{1.5cm}<{\centering}|p{1.5cm}<{\centering}|p{1.5cm}<{\centering}|p{1.5cm}<{\centering}|p{1.5cm}<{\centering}|p{2.5cm}<{\centering}|}
        \hline
		\multirow{1}{*}{} & \multicolumn{7}{c|}{Signal(pb)}&\multicolumn{1}{c|}{Background(pb)}\\
		\cline{2-9}
        \hline
		$M_{W'}$(TeV)& 1 &2 &3 & 4 & 5 &6 &7 &0 \\
		\hline
		Basic cuts	& 18.419& 1.366 &0.281	&0.0902  &0.0372 &0.0180 &0.00976 & 323.988\\
		\hline
		Cut-1  &   10.660  &0.807 &0.165 &0.0531  &0.0219  &0.0107 &0.00576 &47.995 \\
		\hline
		Cut-2 	&4.984	&0.386 &0.0782	&0.0256	&0.0103	&0.00509 &0.00275 &10.036\\
		\hline
		Cut-3     &1.937	&0.154 &0.0311	&0.0101&0.00403 &0.00197 &0.00108 &2.957\\
		\hline
	\end{tabular}
\end{table}

\begin{table}[H]
	\centering
	\caption{\label{Tab:signal2}Same as Table~\ref{Tab:signal1} but for the $\sqrt{s} = 3.46\;{\rm TeV}$ FCC-he. }
	\label{table4}	 \begin{tabular}{|p{2cm}<{\centering}|p{1.5cm}<{\centering}|p{1.5cm}<{\centering}|p{1.5cm}<{\centering}|p{1.5cm}<{\centering}|p{1.5cm}<{\centering}|p{1.5cm}<{\centering}|p{1.5cm}<{\centering}|p{2.5cm}<{\centering}|}
        \hline
		\multirow{1}{*}{} & \multicolumn{7}{c|}{Signal(pb)}&\multicolumn{1}{c|}{Background(pb)}\\
		\cline{2-9}
        \hline
		$M_{W'}$(TeV)& 1 &2 &3 & 4 & 5 &6 &7 &0 \\
		\hline
		Basic cuts	& 55.617&4.687 &1.111	&0.379  &0.159 &0.0797 &0.0431 &483.945 \\
		\hline
		Cut-1  & 30.338    & 2.098&0.507 &0.173  & 0.0731 & 0.0362&0.0197 &17.213 \\
		\hline
		Cut-2 	&22.150	&1.353 &0.337	&0.114	&0.0477	&0.0239 & 0.0131&5.101\\
		\hline
		Cut-3     &20.746	& 1.263&0.311	&0.1065& 0.0442&0.0221 &0.0121 &4.262\\
		\hline
	\end{tabular}
\end{table}

To accurately obtain the sensitivities of the LHeC and FCC-he to the CC NSI parameters that can be estimated with the help of statistical significance ($\mathcal{S}_{stat}$), defined as
\begin{equation}
\mathcal{S}_{stat} = \sqrt{\mathcal{L}}\times \left( \left|\sigma\left(\epsilon^{Y}\right)-\sigma_{SM}\right|/\sqrt{\sigma(\epsilon^{Y})} \right).
\label{eq.antsaz}
\end{equation}
Where $\sigma(\epsilon^{Y})$ is the total cross-section including the $W'$ contributions after cuts applied.
In Fig.~\ref{fig:2} we plot the sensitivities of the $1.30\;{\rm TeV}$ LHeC with $\mathcal{L} = 10\;{\rm fb}^{-1}$ to the CC NSI parameters detected at the 2$\sigma$, 3$\sigma$, and 5$\sigma$ levels for different $M_{W'}$, where the electron beam polarization is taken as $P(e^{-}) = 0$ and $-90\%$ in Fig.~\ref{fig:2}~(a) and (b), respectively.
As can be seen that detectable lower limits are close to each other because the interference terms $\alpha_{int}$ are on the same order of magnitude for different $M_{W'}$. It implies that when the interference effect dominates, the lower limits are insensitive to $M_{W'}$ for $\epsilon^{L}$, while the sensitivities are not symmetric. However, it is not this case for the parameter $\epsilon^{R}$. Because there is no interference effect, $\epsilon^{R}$ is sensitive with $M_{W'}$ and the sensitivities are symmetric. The absolute values of the lower limits on $\epsilon^{R}$ are smaller than those for $\epsilon^{L}$ about two orders of magnitude. Thus we do not show the numerical results about $\epsilon^{R}$ in Fig.~\ref{fig:2}. Furthermore, the electron polarization can make the absolute values of the lower limits smaller. For the electron polarization  $P(e^{-}) = -90\%$, the specifically lower limits on the CC NSI parameter $\epsilon^{L}$ at the $1.30\;{\rm TeV}$ LHeC are  $|\epsilon^{L}| \geq 2.60\times 10^{-4} $, $|\epsilon^{L}| \geq 3.85\times 10^{-4} $ and $|\epsilon^{L}| \geq 6.61\times 10^{-4} $ at the 2$\sigma$, 3$\sigma$, and 5$\sigma$ levels, respectively, as shown in Fig.~\ref{fig:2}~(b).

\begin{figure}[H]
	\centering
	\subfigure[]{	
\centering	
			\begin{minipage}[t]{0.49\linewidth}
			\includegraphics [width=7.5cm] {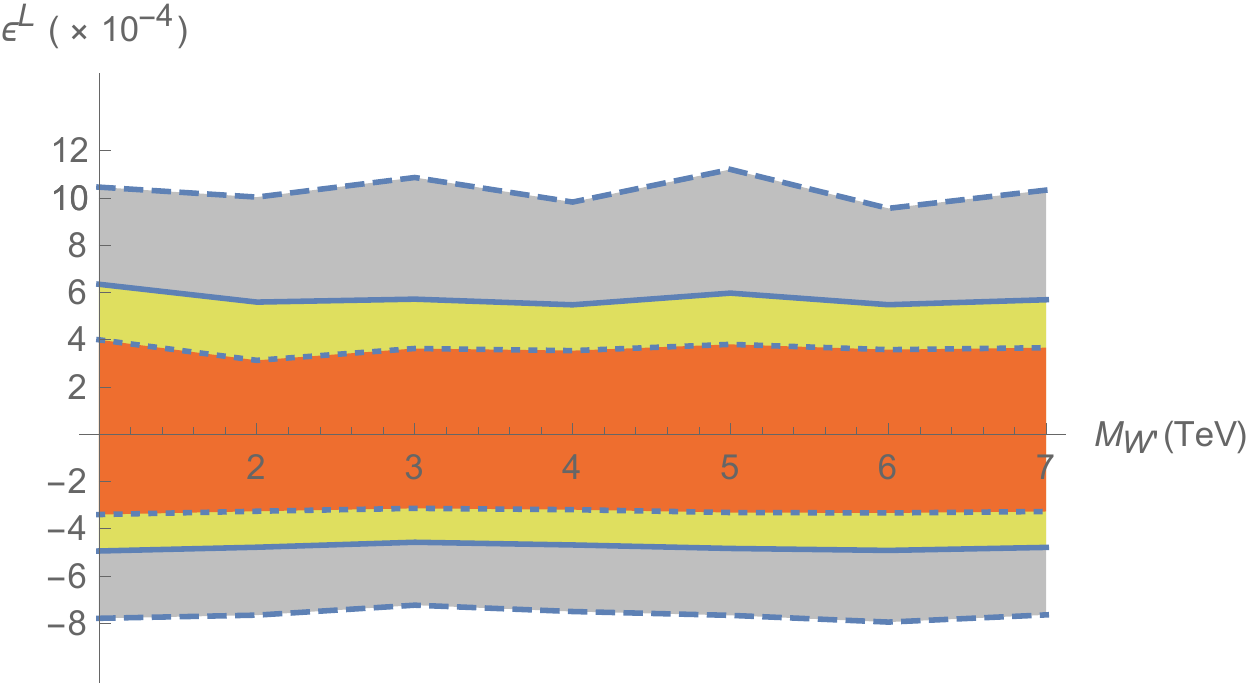}
		\end{minipage}%
	}%
\subfigure[]{	
\begin{minipage}[t]{0.49\linewidth}
	\includegraphics [width=7.5cm] {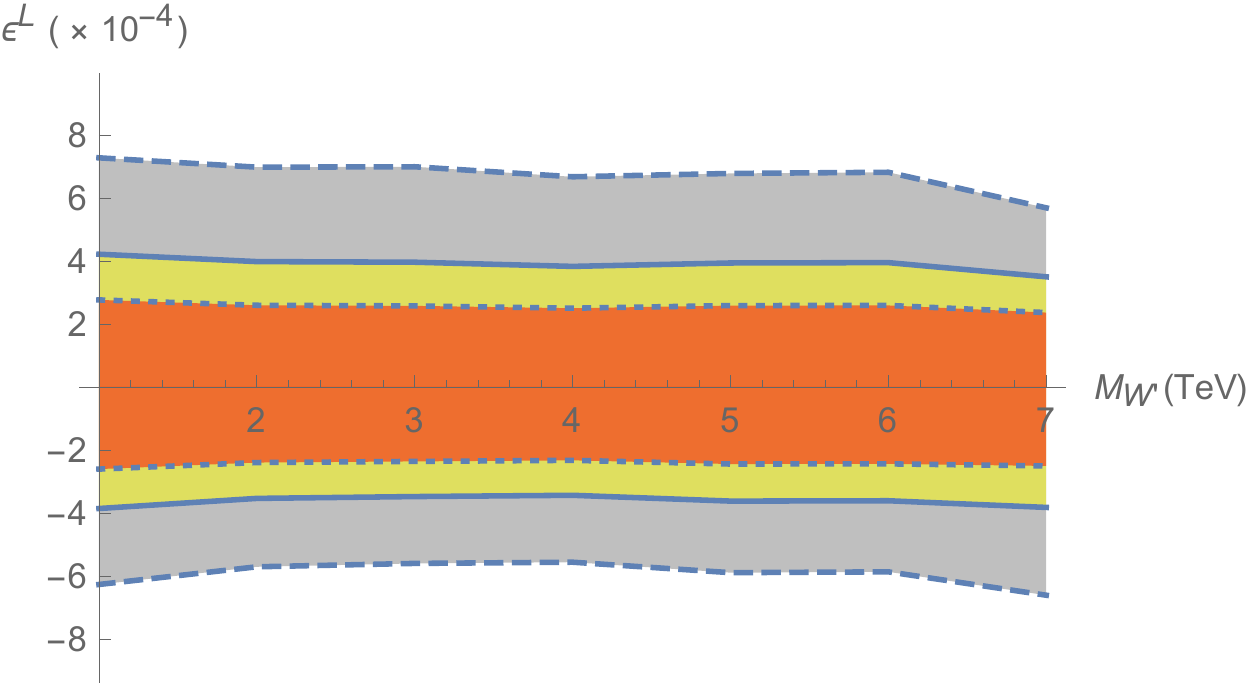}
\end{minipage}%
}%
\caption{\label{fig:2}For $\mathcal{S}_{stat}$ equalling to 2 (dotted line), 3 (solid line) and 5 (dashed line), the sensitivities of the $ 1.30 \;{\rm TeV}$ LHeC with $\mathcal{L}$ = $10\;{\rm fb}^{-1}$ to the CC NSI parameter $\epsilon^{L}$ as functions of $M_{W'}$ for the electron polarization  $P(e^{-}) =0 $ (a) and $ -90\%$ (b).}
\label{fig:2}
\end{figure}

It is obvious that when the integrated luminosity is larger, the $\epsilon^{L}$ is tighter and more sensitive to the detection of new physical signals. We use the similar method to give the lower limits of the sensitivities of the $\sqrt{s} = 1.30\;{\rm TeV}$ LHeC with $\mathcal{L} = 100\;{\rm fb}^{-1}$ and  $P(e^{-}) = -90\%$ to the CC NSI parameters, which are $|\epsilon^{L}| \geq 7.52\times 10^{-5} $, $|\epsilon^{L}| \geq 9.89\times 10^{-5} $ and $|\epsilon^{L}| \geq 1.52\times 10^{-4} $ at the 2$\sigma$, 3$\sigma$, and 5$\sigma$ levels, respectively.

\begin{figure}[H]
\begin{center}
\includegraphics[width=0.6\textwidth]{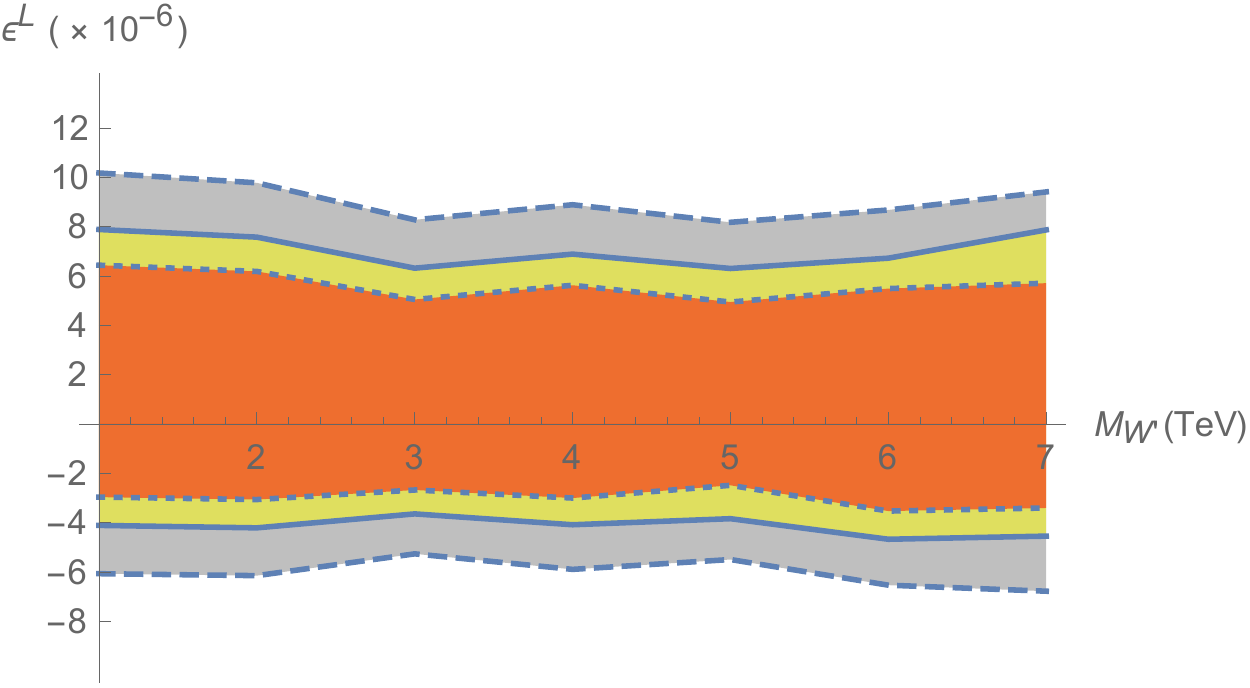}
\caption{Same as Fig.~\ref{fig:2} but for the $ 3.46 \;{\rm TeV}$ FCC-he with $\mathcal{L}$ = $100\;{\rm fb}^{-1}$. }
\label{fig:3}
\end{center}
\end{figure}

As a comparison, we further calculate that the sensitivities of the FCC-he with $\sqrt{s} = 3.46\;{\rm TeV}$, $\mathcal{L} = 100\;{\rm fb}^{-1}$ and  $P(e^{-}) = -90\%$ to the CC NSI parameters, there are $|\epsilon^{L}| \geq 3.56\times 10^{-6} $, $|\epsilon^{L}| \geq 4.67\times 10^{-6} $ and $|\epsilon^{L}| \geq 6.78\times 10^{-6}$ at the 2$\sigma$, 3$\sigma$, and 5$\sigma$ levels, respectively. Details are shown in Fig.~\ref{fig:3}. It is obvious that the absolute values of the lower limits on the CC NSI parameters at the FCC-he are smaller than the results given at the LHeC about two orders of magnitude. All of the numerical results about the parameters $\epsilon^{Y}$ derived in this paper  satisfy the unitarity bounds and the $W'$-decay constraints mentioned in Sec.~\ref{level2}.

Ref.~\cite{Yue:2021snv} has shown that the $13\;{\rm TeV}$ LHC with $\mathcal{L} = 139\;{\rm fb}^{-1}$ can give the expected constraints on the CC NSI parameters, which are $ -1.78\times 10^{-4} \leq \epsilon^{L} \leq 3.36 \times 10^{-4}$ , $ -2.45\times 10^{-4} \leq \epsilon^{L} \leq 4.10 \times 10^{-4}$ and $ -4.08\times 10^{-4} \leq \epsilon^{L} \leq 5.24 \times 10^{-4}$ at the 2$\sigma$, 3$\sigma$, and 5$\sigma$ levels, respectively.
It is obvious that the lower limits on the CC NSI parameter $\epsilon^{L}$ obtained at the $1.30\;{\rm TeV}$ LHeC with $\mathcal{L} = 10\;{\rm fb}^{-1}$ is excluded by the LHC expected constraints due to the small integrated luminosity. However, when the integrated luminosity is increased to $100\;{\rm fb}^{-1}$, our results are not excluded. Besides, the LHeC with $\mathcal{L} = 100\;{\rm fb}^{-1}$ is a little more sensitive to the CC NSI parameters than that of the $14\;{\rm TeV}$ LHC with $\mathcal{L} = 300\;{\rm fb}^{-1}$, and the absolute values of the lower limits on the CC NSI parameters at the FCC-he are smaller than those of the HL-LHC about two orders of magnitude. So we can say that, as a complement to the $pp$ collider,  the LHeC and FCC-he can offer exciting  prospects for detecting the CC NSI effects.

\section{\label{level5} Conclusions }
The LHeC and FCC-he will allow the collisions of electron beams with protons or heavy ions at the HL-LHC. With the c.m. energy greater than ${\rm TeV}$ and a very high luminosity, the LHeC and FCC-he are the new generation of colliders for DIS and complement to $pp$ and $e^{+}e^{-}$ colliders, which also provide powerful opportunities to discover direct or indirect evidence of physics beyond the SM.

A charged gauge boson, $W'$, appears in various extensions of the SM, which can induce the CC NSI. In this paper we focus on the sensitivities of the LHeC and FCC-he to the CC NSI parameters via the subprocess $e^{-}q \to W' \to v_{e}q'$. Under the premise of considering the theoretical constraints on the CC NSI parameters from perturbative unitarity and $W'$ decays, we performed numerical calculations and phenomenological analysis of the signals and the associated backgrounds, and obtained detectable lower limits on the parameter $\epsilon^{L}$ at the LHeC and FCC-he.

Our numerical results show that the lower limits on the parameter $\epsilon^{L}$ obtained at the LHeC and FCC-he are not sensitive to the mass of  $W'$. The $1.30\;{\rm TeV}$ LHeC with $\mathcal{L} = 100\;{\rm fb}^{-1}$ can bring a strong sensitivity to the CC NSI parameter $\epsilon^{L}$, and  the $3.46\;{\rm TeV}$ FCC-he with $\mathcal{L} = 100\;{\rm fb}^{-1}$ can further enhance the sensitivity to $\epsilon^{L}$ by about two orders of magnitude. Specifically, the FCC-he with $\sqrt{s} = 3.46\;{\rm TeV}$, $\mathcal{L} = 100\;{\rm fb}^{-1}$ and  $P(e^{-}) = -90\%$ might detect the CC NSI effects as long as $|\epsilon^{L}| \geq 4.07\times 10^{-6} $, $|\epsilon^{L}| \geq 5.48\times 10^{-6} $ and $|\epsilon^{L}| \geq 6.78\times 10^{-6}$ at the 2$\sigma$, 3$\sigma$, and 5$\sigma$ levels, respectively, which are smaller than those of the HL-LHC about two orders of magnitude.
In summary, studying NSI at the $e^{-}p$ colliders is a promising area for further research, and the $e^{-}p$ colliders have clean environment that will also benefit the physics program at the LHC to explore new physics beyond the SM.

\section*{ACKNOWLEDGMENT}
This work was partially supported by the National Natural Science Foundation of China under Grant Nos. 11875157 and 12147214.

\bibliography{paper}

\end{document}